\documentclass[article]{aa}  

\usepackage{graphicx}
\usepackage{txfonts}
\usepackage{cleveref}
\newcommand{\teff}{\rm T_{eff}}
\newcommand{\logg}{\log(g)}
\newcommand{\feh}{[Fe/H]}
\newcommand{\micro}{\xi_{\rm micro}}
\newcommand{\macro}{\xi_{\rm macro}}

\newcommand{\aboo}{\alpha\: \rm Boo}
\newcommand{\muleo}{\mu\: \rm Leo}
\newcommand{\epsvir}{\epsilon\: \rm Vir}
\newcommand{\betagem}{\beta\: \rm Gem}
\newcommand{\micron}{\mu}
\newcommand{\loggf}{\log(gf)}
\newcommand{\exc}{E_{\rm low}}

\begin{document}

   \title{Chemical Evolution of Ytterbium in the Galactic Disk}

   \author{M. Montelius
          \inst{1,2} \fnmsep 
          \and
          R. Forsberg\inst{2}
          \and 
          N. Ryde\inst{2}
          \and 
          H. J\"onsson\inst{3}
          \and
          M. Af{\c{s}}ar\inst{4,8}
           \and
          A. Johansen\inst{2,5}
          \and
          K. F. Kaplan\inst{6} 
          \and
          H. Kim\inst{7}
         \and 
         \newline
          G. Mace\inst{8}
          \and
          C. Sneden\inst{8}
          \and
          B. Thorsbro\inst{2}
          }

   \institute{Kapteyn Astronomical Institute, University of Groningen, Landleven 12, NL-9747 AD Groningen, the Netherlands     
              \\\email{montelius@astro.rug.nl}
          \and
              Lund Observatory, Department of Astronomy and Theoretical Physics, Lund University, Box 43, SE-22100 Lund, Sweden. 
         \and
             Materials Science and Applied Mathematics, Malm\"o University, SE-205 06 Malm\"o, Sweden 
         \and 
            Department of Astronomy and Space Sciences, Ege University, 35100 Bornova, \.{I}Izmir, Turkey.
        \and 
            Center for Star and Planet Formation, GLOBE Institute University of Copenhagen, Øster Voldgade 5-7, 1350 Copenhagen, \newline Denmark 
        \and
            SOFIA Science Center - USRA, NASA Ames Research Center, Moffett Field, CA 94035, USA 
        \and
            Gemini Observatory/NOIRLab, Casilla 603, La Serena, Chile  
         \and
            Department of Astronomy and McDonald Observatory, The University of Texas, Austin, TX 78712, USA 
            }
    
   \date{Received January 18, 2022; accepted January 31, 2022}


  \abstract
   {Measuring the abundances of neutron-capture elements in Galactic disk stars is an important part of understanding key stellar and galactic processes. In the optical wavelength regime a number of different neutron-capture elements have been measured, however from the infrared H-band only the s-process dominated element cerium has been accurately measured for a large sample of disk stars. The more r-process dominated element ytterbium has only been measured in a small subset of stars so far.}
   {In this study we aim to measure the ytterbium (Yb) abundance of local disk giants using the Yb\,II line at $\lambda_\text{air} = 16498\,$Å. We also compare the resulting abundance trend with cerium and europium abundances for the same stars to analyse the s- and r-process contributions.}
   {We analyse 30 K-giants with high-resolution H-band spectra using spectral synthesis. The very same stars have already been analysed using high-resolution optical spectra using the same method, but the abundance of Yb was not possible to determine from those spectra due to blending issues for stars with \feh $> -1$. In this present analysis, we utilise the stellar parameters determined from the optical analysis.}
   {We determined the Yb abundances with an estimated uncertainty for [Yb/Fe] of 0.1\,dex. From comparison, the trend of [Yb/Fe] follows closely the [Eu/Fe] trend and has clear s-process enrichment in identified s-rich stars. From the comparison, both the validity of the Yb abundances are ensured, and the theoretical prediction of a roughly 40/60 s-/r-process contribution to Yb's origin is supported.
   }
   {These results show that with a careful and detailed analysis of infrared spectra, reliable Yb abundances can be derived for a wider sample of cooler giants in the range $-1.1 < $ \feh $ < 0.3$. This is promising for further studies of the production of Yb and for the r-process channel, key for Galactochemical evolution, in the infrared.}


\maketitle

%

\section{Introduction} \label{sec:intro}
Stellar elemental abundances play a key role in deciphering Galactic chemical evolution. The photosphere of stars carries a chemical imprint of the molecular cloud from which they formed, making them time stamps of the Galactic environment at that epoch. By measuring the stellar abundances from stars originating from different Galactic epochs, the chemical enrichment in the Galaxy can be traced \citep[see e.g.][and references  therein]{bland-hawthorn2016ARA&A..54..529B}. 
The field of Galactic research is increasingly moving to the infrared spectral range, with a different set of spectral lines compared to those available in the optical range. With this comes the possibility of expanding the study of the chemical evolution of the galaxy, in particular to dust covered regions. Different elements can give us different angles on Galactic formation and evolution, and in this study we consider the heavy element ytterbium. 

The production of elements through fusion becomes endothermic above iron which means that the remaining two-thirds of the elements in the periodic table need another production channel. The idea that heavier elements are created by neutron-capture processes was introduced by \citet{burbidge:1957}, giving these elements the name neutron-capture elements. 

As an atomic nucleus captures a neutron, a heavier isotope of that element is created. If this isotope is unstable, a neutron will convert to a proton through $\beta^{-}$-decay and create a heavier element. If the isotope is stable, it can capture further neutrons until eventually $\beta^{-}$-decaying. This outlines two processes, one where the $\beta^{-}$-decay is \textit{slower} than the neutron-capture, and one where the $\beta^{-}$-decay is more \textit{rapid} than the neutron-capture. These are called the s- and r-processes, and together they produce the heavier elements. Indeed, they produce a majority of the isotopes for elements with Z $>$ 30. An example of elements formed through neutron-capture processes are the lanthanides, some of which are studied in this paper: cerium (Ce, Z = 58), europium (Eu, Z = 63), and ytterbium (Yb, Z = 70).

The neutron-capture rate is not constant but depends on both {\it (i)} the local neutron density and {\it (ii)} the neutron absorption cross section of the isotope. This allows some constraints to be put on the environment where they occur. The s-process requires a neutron density of $\leq 10^{11} \textrm{cm}^{-3}$ \citep{busso:1999} and has been shown to take place in the interior of low-mass AGB-stars \citep[see][and references therein]{karakas:2014PASA...31...30K}. The r-process requires a neutron density of around $10^{24}-10^{28} \textrm{ cm}^{-3}$ \citep[see e.g.][]{karlludwig:2007}, which confines it to extremely neutron-dense environments, such as various supernovae (SNe) and the merger of neutron stars. The first observed neutron star merger GW170817 \citep{2017PhRvL.119p1101A} indeed contained traces of r-process element production \citep{tanvir2017ApJ...848L..27T, drout2017Sci...358.1570D}. In contrast, \citet{kobayashi:2020} find from comparing theoretical models to observations that core-collapse SNe, especially magneto-rotational-supernovae (MRSNe), may be the predominant source of r-process production for most neutron-capture elements \citep[see Figure 39 in][]{kobayashi:2020}.

Most neutron-capture elements are produced by a combination of the s- and the r-processes. In the case of Ce the s-process contributed approximately 85\% and the r-process contributed the remaining 15 \% for stars with solar metallicities \citep{birsterzo:2014,prantzos2020MNRAS.491.1832P}. As such, it is usually referred to as 'an s-process element', since the dominating production channel is from this process. As a comparison Eu is referred to as 'an r-process element', with only a 5 \% contribution from the s-process, and the remaining from the r-process \citep{prantzos2020MNRAS.491.1832P}. 


Yb\footnote{The name ytterbium originates from the small Ytterby mine north-east of Stockholm, Sweden, and the element is one of seven discovered from the minerals found there during the 18th and 19th centuries.}, which is the focus of this study, is the second heaviest of the lanthanides and lies at the very end of the series in the periodic table. As such, it is a heavy neutron-capture element which lies in between Ce and Eu with approximately a 40/60 to 50/50 predicted contribution from the s- and the r-processes, respectively \citep{birsterzo:2014,kobayashi:2020,prantzos2020MNRAS.491.1832P}. Nonetheless, the precise nature of its cosmic origin and abundance trend in the Galactic disk has not been well constrained, mainly due to the sparsely available observed abundance data.

Spectroscopically, Yb is, however, relatively well studied, partially due to the interest in studying Ap stars \citep{Cowley:1984} and in ion traps \citep{olmschenk:2009}. The Yb\,II ion has a hydrogenic electron configuration, with an unpaired 6s valence electron. A consequence of this is that the strongest Yb\,II lines are the resonance lines, located at 3289.4\,$\AA$ and 3694.2\,$\AA$\footnote{All wavelengths in this study are wavelengths in air.}, while the other spectral lines are significantly weaker. While both resonance lines have issues with blending, the line at 3694.2\,$\AA$ has been preferred for abundance determination and has been used in a number of studies of low metallicity stars, such as \citet{honda2004ApJ...607..474H,johnson2002ApJS..139..219J,francois:2007A&A...476..935F,roederera2014AJ....147..136R, roedererb2014ApJ...791...32R,sneden:2009ApJS..182...80S}. The analysis is more difficult for stars with higher metallicities, due to a larger impact of blending lines and the difficulty of defining the continuum of the spectrum in the bluer part of the optical regime. As a consequence, these lines have not been used for stars with metallicity{\footnote{The metallicity and other abundances are defined using the notation [A/B] = $\log (N_\mathrm{A}/N_\mathrm{B})_* - \log (N_\mathrm{A}/N_\mathrm{B})_\odot$, where $N_\mathrm{A}$ and $N_\mathrm{B}$ are the number densities of the elements A and B, respectively. The solar abundances used in this study are from \citet{grevesse2007SSRv..130..105G}, unless another reference is given.}} of [Fe/H] $\gtrsim$ -1.5. 

Attempts have been made to fill this gap in metallicity by using the infrared Yb\,II line in the H band at 16498\,$\AA$. This line is the strongest of the infrared Yb lines and was first measured in the laboratory by \citet{humphreys:1959}. The APOGEE survey \citep[$R=\lambda/\Delta\lambda\sim22500$;][]{2017AJ....154...94M} has investigated the utility of this line in their spectra \citep{2021AJ....161..254S}, but do not consider the determined stellar Yb abundances reliable enough to be included in the final data release (DR17, \citet{2021arXiv211202026A}, J. Holtzman et al. in prep.). The reanalysis of APOGEE spectra in the Kepler field by \citet{hawkins:2016} attempts to use the same line, deriving upper limits for the Yb abundance, but do not publish abundances or further analyses. 

In the studies of \citet{bocektopcu2019MNRAS.485.4625B,bocektopcu2020MNRAS.491..544B} they manage to determine Yb for 11 and 10 stars in the open clusters NGC 6940 and NGC 752, respectively, by using the 16498\,$\AA$ Yb\,II line. The temperature of the stars in these samples are around 5000 K, making a CO-blend in the Yb\,II line close to negligible. Furthermore, Yb abundances of three horizontal-branch stars has been determined in \citet{afsar2018ApJ...865...44A} using the very same line, again for stars of $\teff \sim$ 5100 K. 

In this article we also attempt to use the Yb\,II line at 16498\,$\AA$, and we succeed in deriving reliable Yb abundances for 30 K-giants in the Milky Way disk, using high-resolution infrared spectra in the H-band observed with the Immersion Grating \mbox{INfrared} Spectrometer (IGRINS). The spectral resolution it provides, $R\sim 45000$, is higher than APOGEE ($R \sim 22500$), which is important since the Yb\,II line is both weak and heavily blended with a CO molecular line. We investigate the s- and r-process contributions to the cosmic budget of Yb by comparing to the s-element Ce and the r-element Eu, whose abundances have been determined for the very same stars. The Ce abundances are determined from the H-band spectra, while the Eu abundances are taken from \citet{forsberg:2019} where they are determined from optical spectra.

We succeed in measuring reliable Yb-abundances from the 16498\,$\AA$ line for our sample of 30 disk stars. This demonstrates the possibility of larger studies of the r-process channel in the infrared H-band, as well as measuring Yb abundances for stars with metallicities of [Fe/H] $\gtrsim 1$\,dex. Knowing how to properly determine Yb-abundances from the 16498\,$\AA$ line is useful step towards deciphering the formation history and evolution of dust covered regions in the Galaxy, like the Bulge.

In section \ref{sec:obs} we go through the observations and the data used. In section \ref{sec:analysis} we go through the analysis, including uncertainty estimations. In section \ref{sec:res+dis} we present the results and discuss these before concluding in section \ref{sec:conclusion}.

\section{Observations}\label{sec:obs}
For the present work, we have observed 30 K-giants, recording high-resolution spectra in the infrared (IGRINS). These stars were selected from a set of about 500 local disk giants, which have accurate stellar parameters determined through a careful optical spectroscopic analysis (FIES, H. J\"onsson et al. in prep.), see section \cref{sec: stellar param}.

\subsection{IGRINS}
The infrared spectra were observed with the IGRINS spectrograph \citep{igrins,park:2014} on the 4.3-meter Discovery Channel Telescope (DCT; now called the Lowell Discovery Telescope) at Lowell Observatory \citep{mace:2018}, and on the 2.7-meter Harlan J. Smith Telescope at McDonald Observatory \citep{mace:2016} on the civil dates listed in Table \ref{tab:obs}. 

\begin{table*}[h!]
\caption{IGRINS observing log, sorted to match Table \ref{tab:data}.}             
\label{tab:obs}      
\centering                          
\begin{tabular}{c c c c c c c}        
\hline\hline                 
Star &  2MASS star name & H$_{\mathrm{2MASS}}$  &  K$_{\mathrm{2MASS}}$  &  Civil & Telescope & Exposure\\
&   & (mag) & (mag) & Date &  & [s]\\
\hline                        
 $\aboo$ &  J14153968+1910558 &   -2.8 & -2.9 & 2015 April 11 & HJST  & 60  \\ 
 $\muleo$ & J09524585+2600248  &    1.3 &  1.2 &  2016 Jan. 31  & HJST  & 26 \\   
 $\epsvir$  &   J13021059+1057329 & 0.9 &  0.8 & 2016 Feb. 2  & HJST  & 26  \\
 $\betagem$ &   J07451891+2801340 & -1.0 & -1.1 & 2016 Jan. 30  & HJST  & 23  \\
 HD102328  &  J11465561+5537416 &   2.9 &  2.6 & 2016 Feb. 2  & HJST  & 32 \\
 HD102328$^{a}$   &  J11465561+5537416 &  2.9 &  2.6 & 2016 Feb. 27 & HJST & 20 \\   
 HIP50583  &   J10195836+1950290 & -0.8 & -0.8 & 2016 June 20 & HJST  & 29  \\ 
 HIP63432  &   J12595500+6635502 & 2.4 &  2.1 & 2016 May 29  & HJST  & 180 \\ 
 HIP72012 &  J14434444+4027333 & 2.6 &  2.4 & 2016 June 16  & HJST  & 120 \\ 
 HIP90344  &  J18255915+6533486 & 2.2 &  2.1 & 2016 June 15 & HJST  & 120   \\ 
 HIP96014 &   J19311935+5018240 & 2.9 &  2.5 & 2016 June 15  & HJST  & 120 \\ 
 HIP102488  &  J20461266+3358128 & 0.2 &  0.1 & 2016 June 19  & HJST  & 26  \\ 
2M17215666  &   J17215666+4301408 &   7.6 &  7.5 & 2016 July 25 & HJST  & 1080   \\ 
KIC3748585  &   J19272877+3848096 &  6.4 &  6.3 & 2016 Nov. 17& DCT & 240  \\ 
 KIC3955590 &  J19272677+3900456 & 7.8 &  7.7 & 2016 Nov. 23  & DCT   & 480 \\ 
 KIC4177025 &  J19434309+3917436 &  7.6 &  7.5 & 2016 Nov. 22  & DCT & 360 \\ 
 KIC5113910  &   J19421943+4016074 & 8.2 &  8.0 & 2016 Nov. 22  & DCT   & 720 \\ 
 KIC5709564  &  J19321853+4058217 &  7.6 &  7.5 & 2016 Nov. 23  & DCT   & 480  \\
 KIC5779724  &   J19123427+4105257 &  8.0 &  7.8 & 2016 Dec. 09  & DCT  & 600   \\ 
 KIC5859492  &  J19021718+4107236 &  7.9 &  7.8 & 2016 Nov. 23  & DCT   & 480 \\
 KIC5900096  &  J19515137+4106378 & 6.0 &  5.8 & 2016 Nov. 22  & DCT  & 120 \\ 
 KIC6465075 & J19512404+4149284  &   8.0 &  7.9 & 2016 Nov. 23  & DCT  & 480 \\
 KIC6547007  &  J19525719+4158129 &  8.3 &  8.2 & 2016 Dec. 9  & DCT   & 600 \\
 KIC6837256 &  J18464309+4223144 &   8.8 &  8.7 & 2016 Nov. 23  & DCT   & 720 \\
 KIC11045542  &  J19530590+4833180 &  8.4 &  8.2 & 2016 Dec. 11 & DCT & 2000  \\ 
KIC11342694 &  J19110062+4906529 &  7.6 &  7.4 & 2016 Nov. 17 & DCT & 480  \\ 
KIC11444313  &   J19014380+4923062 &   9.1 &  9.0 & 2016 Nov. 23  & DCT   & 720\\ 
 KIC11569659 &  J19464387+4934210 &  9.4 &  9.3 & 2016 Dec. 9 & DCT & 2400 \\ 
 KIC11657684 &  J19175551+4946243 &  9.6 &  9.5 & 2016 Dec. 11 & DCT & 3000  \\ 
2M14231899 &  J14231899+0540079 &  8.0 &  7.8 & 2016 June 19 & HJST  & 1200  \\ 
 HD142091$^{a}$  &  J15511394+3539264 &  2.6 &  2.5 & 2016 Feb. 29 & HJST & 44 \\       
\hline                                 
\end{tabular}
\tablefoot{\\DCT: the Discovery Channel Telescope, a 4.3 m telescope at Lowell Observatory, Arizona.\\ 
HJST: the  Harlan J Smith Telescope, a 2.7 m telescope at McDonald Observatory, Texas. \\
$^a$ Data from the IGRINS Spectral Library \citep{park:18} }
\end{table*}

IGRINS provides a spectral resolving power of $R\sim 45000$ spanning the full H and K bands ($1.45-2.5\,\micron$m), recorded in one exposure, even though for this paper we are mainly interested in the Yb\,II line at 16498\,$\AA$. The stars were observed in an ABBA nod sequence along the slit to permit sky background subtraction. We aimed for signal-to-noise ratios (SNR) of at least 100, leading to exposure times for these bright objects in the range of $30$ to $3000$ seconds, see Table \ref{tab:obs}. In conjunction with the science targets, telluric standard stars (typically rapidly rotating, late B to early A dwarfs) were observed at similar air masses. All the spectra were optimally extracted (thus, not a simple sum extraction) using the IGRINS reduction pipeline \citep{igrins_pipeline:2017}. The pipeline extracts wavelength calibrated spectra after flat-field correction and A-B frame subtraction. Obvious cosmic-ray signatures in the spectra, being in emission or very strong one-pixel features in absorption, which are not taken care of by the pipeline were carefully eliminated. 

In order to divide out the telluric lines, the observed spectra were divided by the spectra of the telluric standard stars. Every order of the divided spectra is continuum normalised with the IRAF task {\tt continuum} \citep{IRAF} and were then combined with the task {\tt scombine} allowing the addition of overlapping regions of subsequent orders. The parts of the edges of the overlapping regions with no traceable continuum and spurious edge effects were omitted. This resulted in one normalised and stitched spectrum across the entire infrared H and K bands for each star. One of the stars, HD102328, was observed twice and the two reduced spectra are analysed independently.

\subsection{FIES}\label{sec: FIES data}
As already mentioned, the 30 stars in our infrared IGRINS sample is a sub-sample of some 500 local disk giants. This larger sample of stars were observed with the high-resolution spectrograph FIES \citep{fies:telting:2014}, mounted on the Nordic Optical Telescope (NOT) at La Palma, Spain. FIES has a resolution of $R\sim 67 000$, with a wavelength range of 3700-8300\,$\AA$. The SNR of these spectra is generally high, often around 200. More about these spectra and observations can be found in \citet{jonsson2017a}.


\section{Analysis}\label{sec:analysis}
In this section we will go through the stellar parameters determined from optical spectra in Section \ref{sec: stellar param}. Supporting abundances determined from either optical or IGRINS spectra are described in section \ref{sec: other elem abundances}. We cover the method and atomic data used in Section \ref{sec: spectral synthesis} and \ref{sec: atomic data}. Details of the Yb abundance and handling of the CO-blend are described in Section \ref{sec: Yb determination} and lastly the uncertainties are covered in Section \ref{sec:unc}.

\subsection{Stellar parameters}\label{sec: stellar param}
Our sample of stars have accurate stellar parameters determined through a careful spectroscopic analysis of the optical FIES data. This analysis will be described in J\"onsson et al. (in prep.), but builds upon and improves the analysis described in \citet{jonsson2017a}. The main difference between the determination of stellar parameters in \citet{jonsson2017a} and in J\"onsson et al. (in prep), is that the latter use the entire wavelength region available in the FIES spectra, and is not restricted to 5800-6800\,$\AA$ as was done in the former. This of course leads to more available Fe\,I, and in particular, Fe\,II lines improving the accuracy of the determined stellar parameters in general, and also removes the systematic difference of 0.1\,dex in $\logg$ when comparing to asteroseismic values found in the 2017 analysis. 

The uncertainties of the stellar parameters from the optical analysis 
are estimated to typically be 50 $\textrm{K}$ in $\teff$, 0.1\,dex in $\logg$, 0.05\,dex in \feh, and 0.1 km/s in $\micro$.

Macroturbulence, $\macro$, depends on large scale motion in the stellar atmosphere and the instrumental profile of the spectrograph. As such it can not be adopted from the optical data and is instead determined from a set of Fe\,I lines in the infrared with accurate van der Waals broadening parameters supplied by P. Barklem (2018, priv. comm.).

\subsection{Other elemental abundances}\label{sec: other elem abundances}
In this paper we also make use of some of the stellar abundances derived from the optical spectra; in particular C, N and O for modelling the ubiquitous molecular lines in the IGRINS spectra. The C, N and O abundances are derived from the forbidden O\,I doublet at 6300 and 6364$\,\AA$ and selected CN and C$_2$ molecular lines. The general uncertainties of these abundances are estimated to be of the order of 0.05\,dex. To ensure the highest accuracy in modeling the blending CO line, the C abundances for some of the stars have been supplemented with measurements from C\,I lines in the IGRINS spectra. 

In the discussion in section \ref{sec:res+dis}, the star's abundances of the neutron-capture elements Ce and Eu are used. The Ce abundances have been determined using the Ce\,II lines at 16595.18\,$\AA$ and 17058.88\,$\AA$. The two lines have minor blends, Mg\,I and Fe\,I respectively, which do not appear to affect the measurements significantly. The derived abundances agree well with the optical measurements from J\"onsson et al. (in prep.). The Eu abundances are taken from \citet{forsberg:2019}, which uses the same FIES spectra as J\"onsson et al. (in prep.), but with the stellar parameters from \citet{jonsson2017a}.

Within the J\"onsson et al. (in prep.) analysis several stars with unexpectedly high abundances of s-elements were found. These stars have, at the same time, `normal' abundances in the other elements, following the general [X/Fe]-trend versus [Fe/H]. This indicates partly the precision of these optical measurements and partly the high confidence we can have in the determined s-element enhancement. Three of these stars are among the presently analysed sample of stars and stand out in the s-process element Ce (see Figure \ref{fig:n-cap}). This is useful when we discuss our Yb measurements. 

In Table \ref{tab:data} we provide all stellar parameters and abundances for the IGRINS sample of stars, and the s-enriched stars are indicated as well as a classification of the stars into different stellar populations, the thin disk, the thick disk, or the halo. This division into stellar populations is made using a combination of high- and low $\alpha$ abundances ([Mg/Fe]) combined with kinematics in a scheme similar to the method used in \citet{lomaeva:2019,forsberg:2019}.

\subsection{Spectral Synthesis}\label{sec: spectral synthesis}
We determine the abundances using the spectral synthesis code Spectroscopy Made Easy \citep[SME][]{sme,sme_code}. SME interpolates a model atmosphere from a grid of 1D spherically symmetric MARCS models \citep{gustafsson:2008}, computed using Local Thermodynamic Equilibrium (LTE). The abundance is determined by finding a best fit to the best fit to the observed spectrum using $\chi^2$-minimisation of the synthetic spectrum from the model atmosphere grid. 

In order to synthesise a spectrum SME requires, besides a line list with atomic data and a model atmosphere, a segment within which line mask and several continuum masks are defined. We define the line- and continuum-masks manually and examine all synthetic spectra by eye to ensure a good fit to the observed spectra. In the cases where the final spectra still have some modulation in their continuum levels, these are taken care of by defining specific local continua around the spectral line in those stars. 

\subsection{Atomic data}\label{sec: atomic data}
The basis of the linelist used in this study is an extraction of atomic lines from the VALD3 database \citep{rybachikova:2015}. As a number of the lines in the studied wavelength range lack reliable measurements of their log(gf) values, these values have been shifted to match the solar spectra, most notably the Fe\,I and Zn\,I lines in the vicinity of the Yb\,II line. Additionally, a nearby Ni\,I line has had its $\loggf$ value adjusted to fit the spectrum of $\muleo$ using a [Ni/Fe]=0.07, since it presented a blending problem for supersolar metallicity stars while not being visible in the Sun.

For molecular lines, a number of additional sources have been used: $^{12}$C$^{14}$N and $^{13}$C$^{14}$N from \citet{sneden:2014}; $^{12}$C$^{16}$O and $^{13}$C$^{16}$O from \citet{li:2015}; and $^{16}$OH from \citet{brooke:2016}. H$_2$O lines are present in the H-band, but for the stars in this paper the impact on the spectra around the Yb\,II line is below 0.1$\ \%$ of the continuum level. As this is far below the noise level the H$_2$O lines are not included. No other molecules are expected to show significant lines in K-giant H-band spectra. 

For the 16498\,$\AA$ line of Yb\,II used in this study, the atomic data is sourced from \citet{biemont:1998}, which presents $\log$(gf) values calculated using the Hartree-Fock approximation \citep{cowan:1981}. The data for the line is given in Table \ref{tab:linelist} together with the most prominent blends. 

The Ce\,II lines used to determine Ce abundances for the stars have wavelengths and excitation energies from \citet{corliss:1973}, but there are no laboratory measurements of the $\loggf$ values. They have instead been determined astrophysically, by adjusting $\loggf$ line by line such that the mean abundances agree with the measurements of the same stars in J\"onsson et al. (in prep.). 


\begin{table}[h]
\caption{Spectral line data for the line used to determine Yb abundances and nearby molecular lines, also the Ce\,II lines used to measure Ce. The log(gf) values for the Ni\,I and Ce\,II lines are the ones measured in this study. The Ce\,II line at 16595\,$\AA$ has also been measured astrophysically by \citet{cunha:2017}, the $\Delta\loggf=0.076$.}             
\label{tab:linelist}      
\centering                          
\begin{tabular}{l c c c c}        
\hline\hline                 
&$\lambda_\textrm{air}$& $\loggf$ & $\exc$ &  Reference\\    
 & [Å] & & [eV] &  \\
\hline                        
   CN &   16496.815 & -1.584 & 1.307 & S14\\
   OH &   16497.737 & -5.448 & 1.266 & B16\\
   OH &   16497.979 & -5.448 & 1.266 & B16\\
   CO &   16498.273 & -5.606 & 1.847 & L15\\
   Yb\,II & 16498.420 & -0.640 & 3.017 & B98\\
   Ni\,I & 16499.131 & -1.363 & 6.257 & K08\\
   Ce\,II & 16595.180 & -2.114 & 0.122 & C73\\
   Ce\,II & 17058.880 & -1.425 & 0.318 & C73\\
\hline                                   
\end{tabular}
\tablefoot{\\S14: \citet{sneden:2014}, B16: \citet{brooke:2016}, L15: \citet{li:2015}, B98: \citet{biemont:1998}, K08: \citet{K08}, C73: \citet{corliss:1973}.}
\end{table}

\begin{figure*}[h]
\centering
\includegraphics[width=\textwidth]{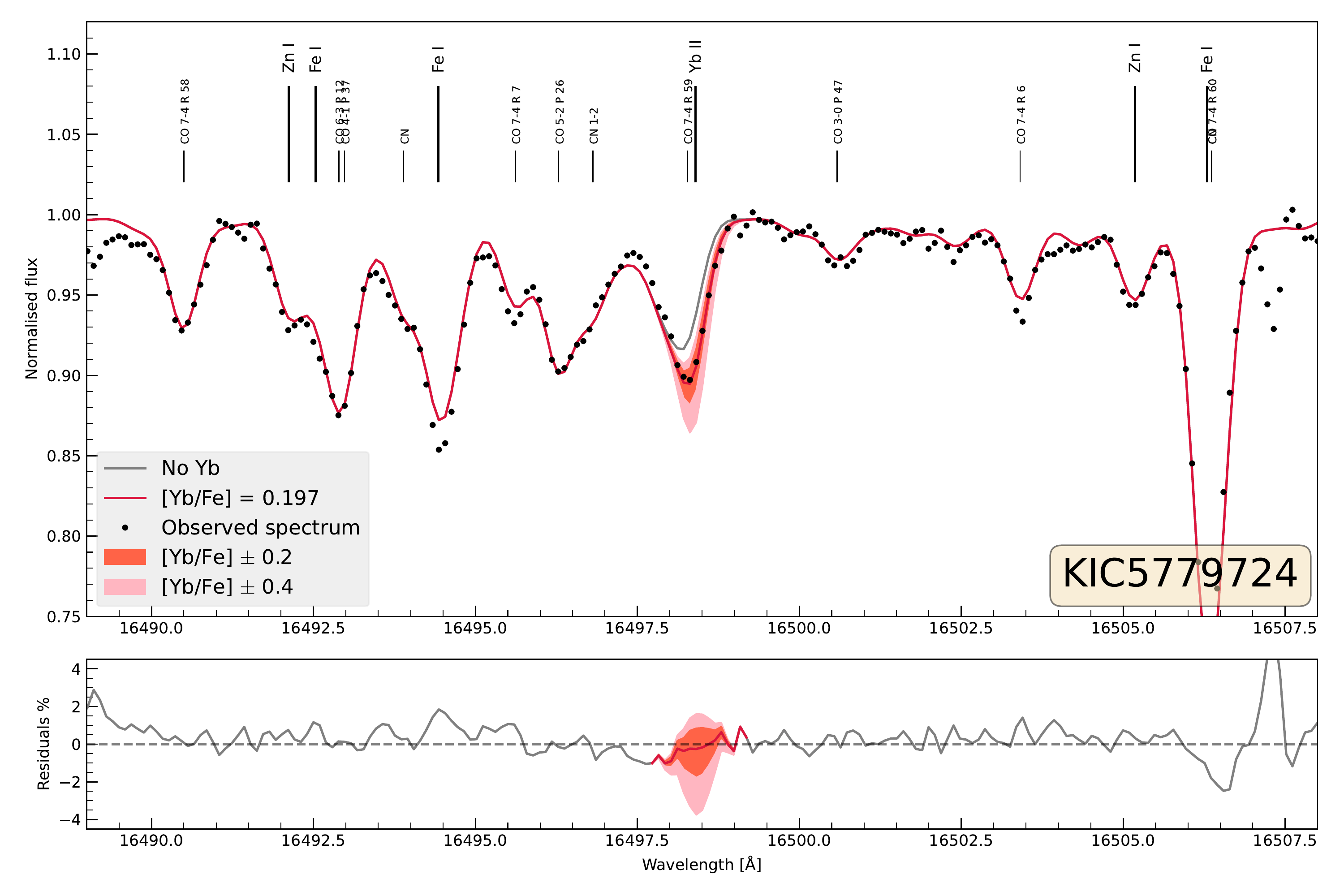}
    \caption{Comparison between the observed and the synthesised spectrum for KIC5779724 around the Yb\,II line at $\lambda_\textrm{air}=16498.42$\,$\AA$, with the residuals plotted below. Additional curves show the impact of increasing or decreasing the Yb abundance by 0.2 and 0.4\,dex respectively.}
    \label{fig:intervalplot}
\end{figure*}

\begin{figure*}[h]
\centering
\includegraphics[width=\textwidth]{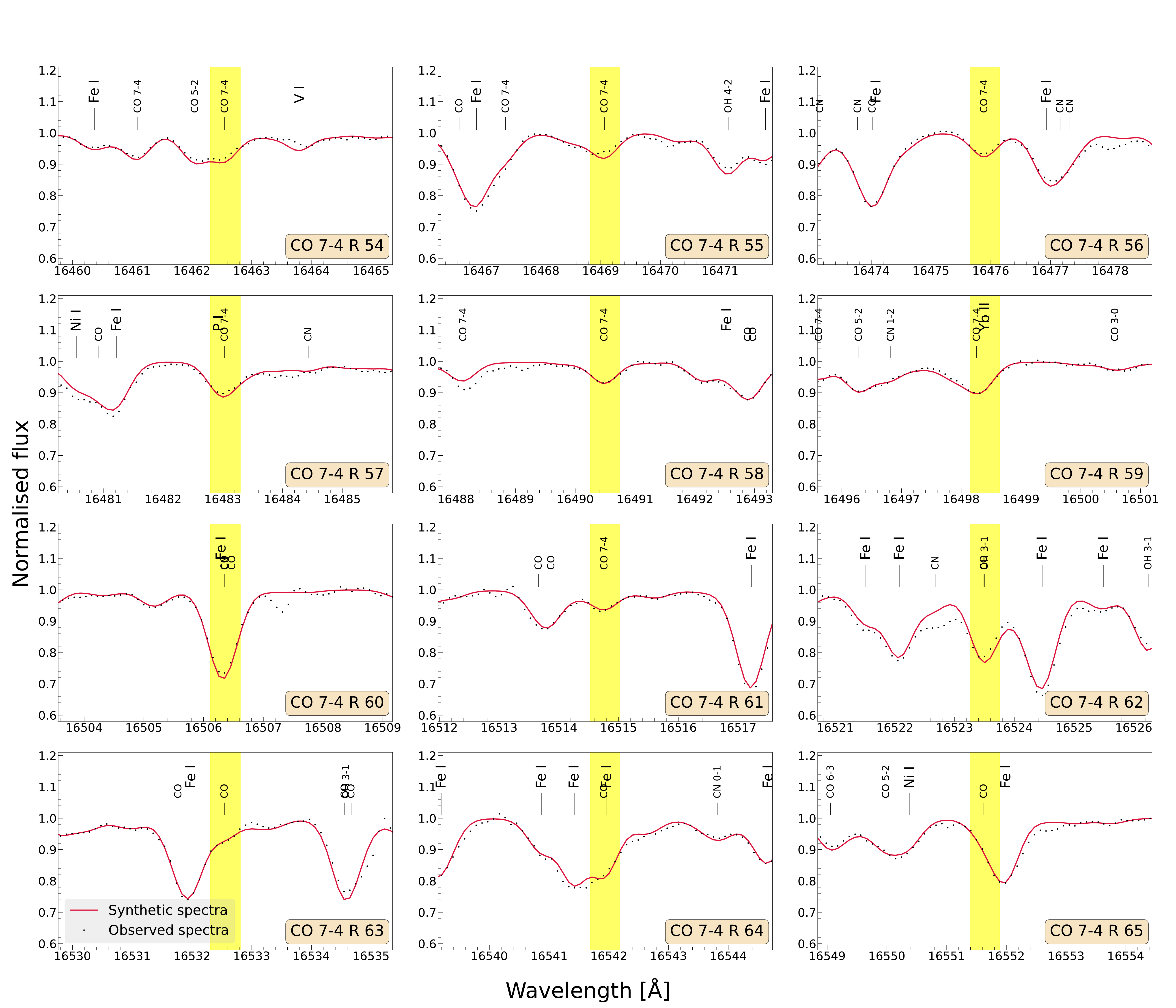}
    \caption{Comparison between the observed and the synthesised spectra for the star KIC5779724 around the CO 7-4 lines with rotational transitions between R 54 and R 65. The line blending the Yb\,II line is from the CO 7-4 R 59 transition. Details of the observation are listed in Table \ref{tab:obs} and the parameters and stellar abundances are listed in Table \ref{tab:data}.}
    \label{fig:COplot}
\end{figure*}

\subsection{Ytterbium determination}\label{sec: Yb determination}
The Yb\,II 16498\,$\AA$ line is rather weak, although detectable in high-resolution spectra with high SNR. The main difficulty in analysing the line comes from the molecular lines which blend it. Primarily, this blend is made up of a CO line, but with a significant contribution from an OH doublet for cooler, low-metallicity stars. 

The strength of the CO blend presents the largest obstacle to measuring the Yb abundance in our stars. To make sure that the CO lines are properly modeled, the fit of nearby CO lines has been inspected. From this inspection, it has become evident that the fit between the observed and the synthesised spectra is dependent not only on the species of the lines, but also on the vibrational and rotational state of the CO molecule. The CO line at 16498\,$\AA$ is caused by a 7-4 vibrational, R 59 rotational transition. Several other lines from this vibrational state can be seen in Figure \ref{fig:intervalplot}. Lines from lower rotational transitions, such as the 7-4 R 6 line at 16503.4\,$\AA$ appear to be too weak, while the more highly excited 7-4 R 58 line at 16490.5\,$\AA$ appears to fit the observed spectra well. The reason for this discrepancy is likely to be related to the atmospheric depth where the lines form: lines with lower excitation energies form further out where the modeling of the atmosphere is less certain and 3D and NLTE effects may play a large role. 

Figure \ref{fig:COplot} shows CO lines from vibrational-rotational transitions similar to the CO line blending the Yb\,II line. As these lines should form in a similar way and the line-to-line strength differences are very accurately understood, the fit between them and the spectra should act as a proxy to evaluating the modeling of the Yb-blending CO line. Because of issues with blending, the quality of the fit is not always clear, although the lack of major discrepancies reassures that the CO line of interest should not spoil the measurement of Yb.

The OH lines are less disruptive for the majority of the stars as the lines are generally weaker and further to the blue of the Yb\,II line. The low excitation energies of the lines make their strengths inversely dependent on temperature, and move their formation further out into the atmosphere, increasing the risk of modeling errors due to, e.g, 3D effects. The strong oxygen enhancement for metal-poor stars further increases their strength relative to other lines, which in combination with a low $\teff$ can make the OH lines equal to the CO line in strength and significantly disrupt the Yb\,II line. The placement of the line mask and the visual inspection of the fit are therefore critical for measuring Yb in metal-poor stars with $\teff < 4300$ K.

A nearby CN line at 16496.8\,$\AA$ may be a blending factor at lower resolutions, but does not influence the Yb\,II line in the spectra studied in this paper.

\subsection{Uncertainty analysis}\label{sec:unc}
\subsubsection{Systematic uncertainties}
The main sources of systematic uncertainty in abundance measurements are errors in stellar parameters, atomic data and assumptions made in the model atmosphere. As described in section \ref{sec: stellar param}, the stellar parameters derived by J\"onsson et al. (in prep.) are believed to be both precise and accurate, based on comparisons to benchmark values for $\teff$ and $\logg$ from angular diameter and asteroseismological measurements. Assessing whether there is a bias in the elemental abundances due to stellar parameters is sometimes done by plotting them against each other. This validation can not be performed for this study, as there is an observational bias in our sample. Low metallicity stars are often further away, and to allow such stars to be observed with the optical 2.5m NOT/FIES telescope/high resolution spectrograph combination, the metal-poor targets tend to have lower surface gravities and temperatures. 

NLTE corrections have not been computed for Yb, and as such their impact can not be quantified. As all stars have similar temperatures and surface gravities ($\teff{_{mean}} = 4520\pm250$ K, $\logg_{\rm mean} = 2.2\pm 0.5$) we would expect possible NLTE effects to have a small impact on the shape of the [Yb/Fe] vs [Fe/H] trend.

There are seven stable isotopes of Yb, two of them with hyperfine structure splitting. Both factors are assessed for the resonance lines of Yb \citep[e.g.][]{martensson:1994}, but no information is available for the 16498\,$\AA$ line. No sign of additional broadening from such factors has been seen in the spectra.

\subsubsection{Random uncertainties}\label{sec:RandUnc}
To estimate the uncertainties in the abundance measurements stemming from random uncertainties in the stellar parameters, a Monte Carlo technique has been used. For each star, the spectra have been reanalysed 500 times using parameters drawn from normal distributions with the adopted values of $\teff$, $\logg$, \feh, $\micro$ and [C/Fe] taken as the mean of the distribution, and the measurement uncertainty used as the standard deviation. The adapted uncertainties are 50 $\textrm{K}$ in $\teff$, 0.1\,dex in $\logg$, 0.05\,dex in \feh, 0.1 km/s in $\micro$ and 0.05\,dex in [C/Fe]. Varying these parameters will give a good measure of the uncertainty in the analysis of the Yb\,II line itself as well as the uncertainty introduced by the blending lines. The mean average deviation of the Yb measurements performed with the varied stellar parameters is then adopted as the uncertainty in [Yb/Fe]. In Figure \ref{fig:errorbar} the upper and lower quantiles are used to create error bars for the measurement of each star. 

\begin{figure}[h]
\centering
\includegraphics[width=10cm]{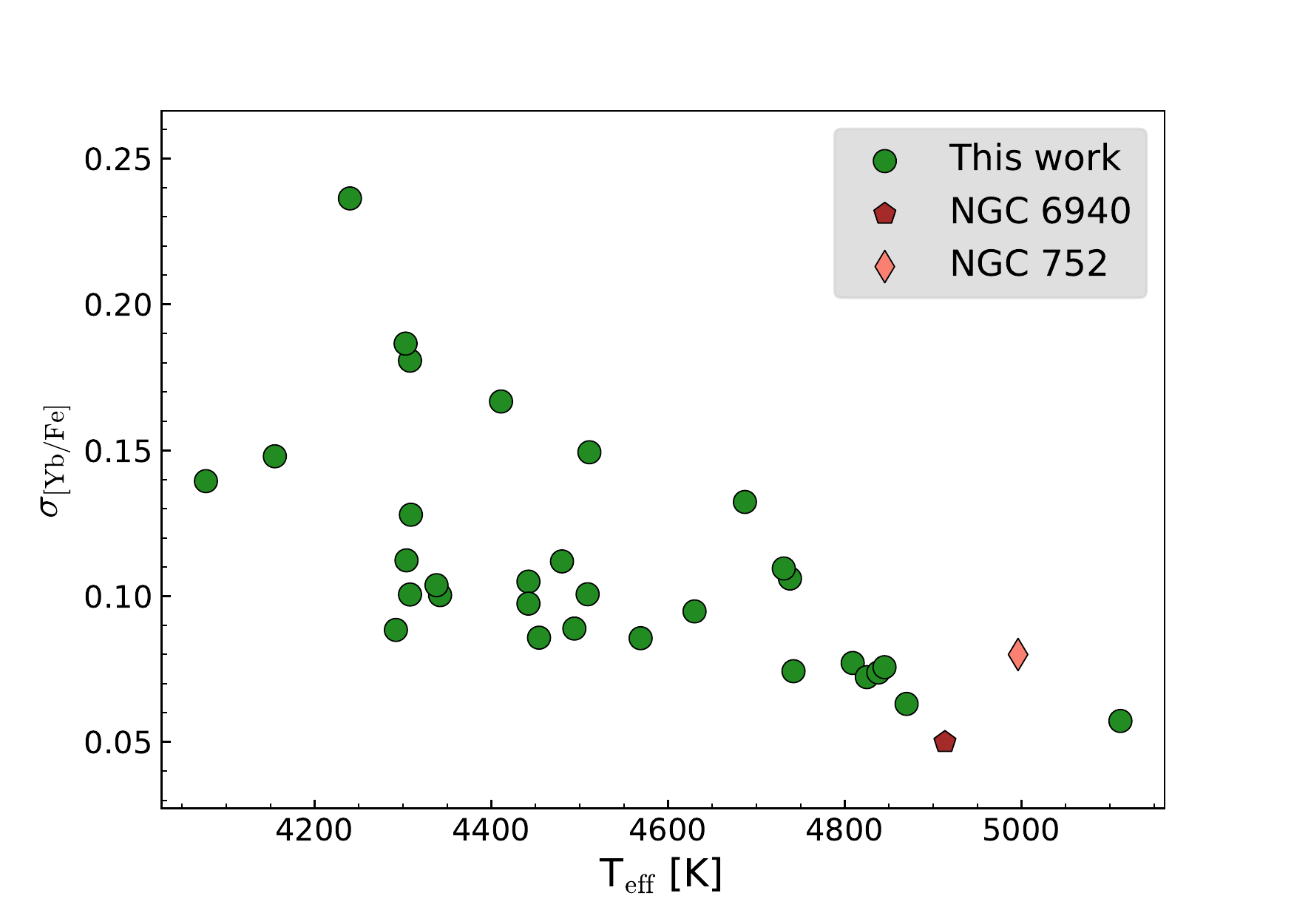}
    \caption{Random uncertainties calculated as described in section \ref{sec:RandUnc} plotted against the $\teff$ of the stars. The mean temperature and spread in Yb abundance for NGC 6940 and NGC 752 \citep{bocektopcu2019MNRAS.485.4625B,bocektopcu2020MNRAS.491..544B} are also marked as a comparison.}
    \label{fig:unc_Teff}
\end{figure}

The mean random uncertainty for our entire sample is 0.11\,dex, which is slightly larger than the spread in Yb abundance measured for the open clusters NGC 6940 and NGC 752 of 0.05-0.08\,dex \citep{bocektopcu2019MNRAS.485.4625B,bocektopcu2020MNRAS.491..544B}. The larger uncertainty in our study is possibly caused by a lower mean temperature in the stars studied here, increasing the influence of the molecular blends. Figure \ref{fig:unc_Teff} supports this view, showing the calculated uncertainties for Yb plotted against $\teff$. A clear trend can be seen with large uncertainties for cooler stars and smaller values for the hotter stars, matching the results for the open clusters.

The star HD102328 has been observed twice, allowing us to estimate the effect of random errors unrelated to stellar parameters, such as SNR and continuum determination. The difference in [Yb/Fe] between the two spectra is 0.06\,dex, matching the spread of 0.05-0.08 from \citet{bocektopcu2019MNRAS.485.4625B,bocektopcu2020MNRAS.491..544B} well. Note that the difference in SNR between the spectra are sizeable (125 and 250) and likely contributes a significant part to the difference between the measurements.

\begin{table*}
\caption{Program stars observed with IGRINS. The [Yb/Fe] and [Ce/Fe] abundances are determined in this work, whilst the [Eu/Fe] abundances are the ones from \citet{forsberg:2019}. The distances are from \citet{mcmillan:2018}. The stellar parameters and classification into different stellar populations are determined in J\"onsson et al. (in prep.), see Section \ref{sec: stellar param} - \ref{sec: other elem abundances}. SNR is determined using the code DER\_SNR \citep{DERSNR2008} applied to a set of continuum regions in the H-band.}             
\label{tab:data}      
\centering                          
\begin{tabular}{c c c c c c c c c c c}        
\hline\hline                 
Star & $\teff$ & $\logg$ & \feh & $\micro$ & Distance & Stellar pop. & SNR & [Yb/Fe] & [Eu/Fe]$^a$ & [Ce/Fe] \\    
     & [K]   & [dex] & [dex] & [km/s]   & [pc] &  &  & [dex] & [dex] & [dex] \\
\hline                        
$\aboo$     &  4308  &  1.7 &  -0.56  &  1.8  & $\ldots$ &   thick & 155 & 0.111 & 0.317 & -0.293 \\
$\muleo$   &  4494  &  2.5 &   0.29  &  1.5  & $\ldots$ &   thin & 169 & 0.090 & 0.034 & -0.283 \\
$\epsvir$   &  5112  &  3.0 &   0.15  &  1.6  & $\ldots$ &   thin & 299 & -0.002 & $\ldots$ & 0.012 \\
$\betagem$  &  4809  &  2.8 &   0.00  &  1.4  & $\ldots$ &   thin & 336 & -0.210 & 0.016 & -0.101  \\
HD102328  &  4442  &  2.5 &   0.30  &  1.5 &  70$\pm$  1 &  thin & 250 & 0.068 & -0.027 & -0.218 \\
HD102328$^{a}$&  4442  &  2.5 &   0.28  &  1.5  & 68$\pm$ 1 & thin & 125 & 0.127 & -0.027 & -0.223 \\
HIP50583  &  4292  &  1.7 &  -0.49  &  1.7  & $\ldots$ &  thin & 86 & 0.382 & 0.193 & 0.020 \\
HIP63432  &  4155  &  1.3 &  -0.65  &  1.9  &  199$\pm$ 4 & thick & 275 & 0.317 & 0.292 & -0.039 \\
HIP72012  &  4077  &  1.4 &  -0.19  &  1.55 &  196$\pm$ 3 &  thin & 286 & 0.055 & 0.182 & -0.123\\
HIP90344  &  4454  &  2.2 &  -0.37  &  1.4  &   91$\pm$ 1 &  thin & 283 & 0.403 & 0.223 & 0.013\\
HIP96014  &  4240  &  1.6 &  -0.35  &  1.7  &  139$\pm$ 2 &  thin & 414 & 0.052 & 0.086 & -0.149\\
HIP102488 &  4742  &  2.5 &  -0.16  &  1.6  & $\ldots$ &  thin & 343 & 0.247 & 0.137 & -0.100\\
2M17215666 & 4342  &  1.6 &  -1.10  &  1.7  & 1900$\pm$ 200 & halo, s-rich & 165 & 0.424 & 0.511 & 0.207 \\
KIC3748585 &  4569  &  2.6 &  0.08  &  1.3  &  360$\pm$ 10 & thin & 205 & 0.008 & -0.037 & -0.105\\
KIC3955590 &  4411  &  2.2 &  0.06  &  1.6 &  950$\pm$ 50 & thin & 140 & -0.114 & -0.079 &  -0.232\\
KIC4177025 &  4309  &  1.7 & -0.43  &  1.6  & 1200$\pm$  80 & thick & 185 & 0.263 & 0.607 & -0.254\\
KIC5113910 & 4338  &  1.7 & -0.37  &  1.6  & 1900$\pm$ 200 &  thin & 183 & 0.245 & 0.122 & -0.040\\
KIC5709564 & 4687  &  2.2 & -0.35  & 1.7  &  640$\pm$  20 & thick & 137 & -0.050 & 0.406 & -0.330\\
KIC5779724 & 4303  &  1.6 & -0.48  &  1.7  & 1500$\pm$ 100 & thick & 215 & 0.197 & 0.417 & -0.225\\
KIC5859492 & 4511  &  2.4 &  0.14  &  1.5  &  780$\pm$ 30 & thin & 181 & -0.262 & 0.030 &  -0.288\\
KIC5900096 & 4480  &  2.5 &  0.22  & 1.5  &  313$\pm$  6 &  thin & 179 & -0.047 & 0.071 & -0.210\\
KIC6465075 &  4825  &  2.8 &  -0.28  &  1.3  &  550$\pm$ 20 &  thin & 146 & 0.300 & 0.169 & 0.071\\
KIC6547007 &  4738  &  2.4 &  -0.72  &  1.4  &  750$\pm$ 30 & thick & 281 & 0.201 & 0.397 & -0.149 \\
KIC6837256 & 4731 &  2.3 & -0.68 & 1.5  & 1250$\pm$ 60 &  thick & 148 & 0.209 & 0.306 & -0.226 \\
KIC11045542 & 4304  & 1.6 & -0.52  &  1.5  & 1900$\pm$ 200 & thin & 294 & 0.186 & 0.085 & -0.201\\
KIC11342694 & 4509  & 2.8 &  0.21  &  1.3  &  440$\pm$  10 & thin & 139 & 0.044 & 0.044 & -0.118\\
KIC11444313 & 4630  & 2.3 & -0.09  &  1.5  & 1400$\pm$  80 & thin & 160 & 0.070 & 0.146 & -0.144 \\
KIC11569659 & 4838  & 2.4 & -0.33  &  1.6  & 1510$\pm$  80 & thin & 246 & 0.168 & 0.162 & -0.061 \\
KIC11657684 & 4870  & 2.6 & -0.26  &  1.5  & 1700$\pm$ 100 & thin, s-rich & 184 & 0.448 & 0.096 &  0.351 \\
2M14231899  &  4308  &  1.8 &  -0.77  &  1.6  & 1900$\pm$ 300 &  thick, s-rich & 284 & 0.639 & 0.547 &  0.179 \\
HD142091$^{a}$&  4845  &  3.3 & 0.06  &  1.2 & 30$\pm$ 0.1 & thin & 223 & 0.066 & $\ldots$ & -0.217 \\
\hline                                   
\end{tabular}
\tablefoot{We use A(Fe)$_\odot = 7.45$, A(Yb)$_\odot = 1.08$ \citep{grevesse2007SSRv..130..105G} and A(Ce)$_\odot = 1.58$,  A(Eu)$_\odot = 0.52$ \citep{grevesse:2015}.\\
$^a$ The [Eu/Fe] values in \citet{forsberg:2019} are reported to be systematically too high, possibly originating from systematic uncertainties in $\logg$ as reported in \citet{jonsson2017a}. We lower all these abundances by 0.10\,dex both in this table and in subsequent figures, such that the overall thin disk in [Eu/Fe] goes through the solar value, as seen in Figure \ref{fig:n-cap}. \\ 
$^b$ Data from the IGRINS Spectral Library \citep{park:18} }
\end{table*}

\section{Results and Discussion}\label{sec:res+dis}
\begin{figure}[h]
\centering
\includegraphics[width=10cm]{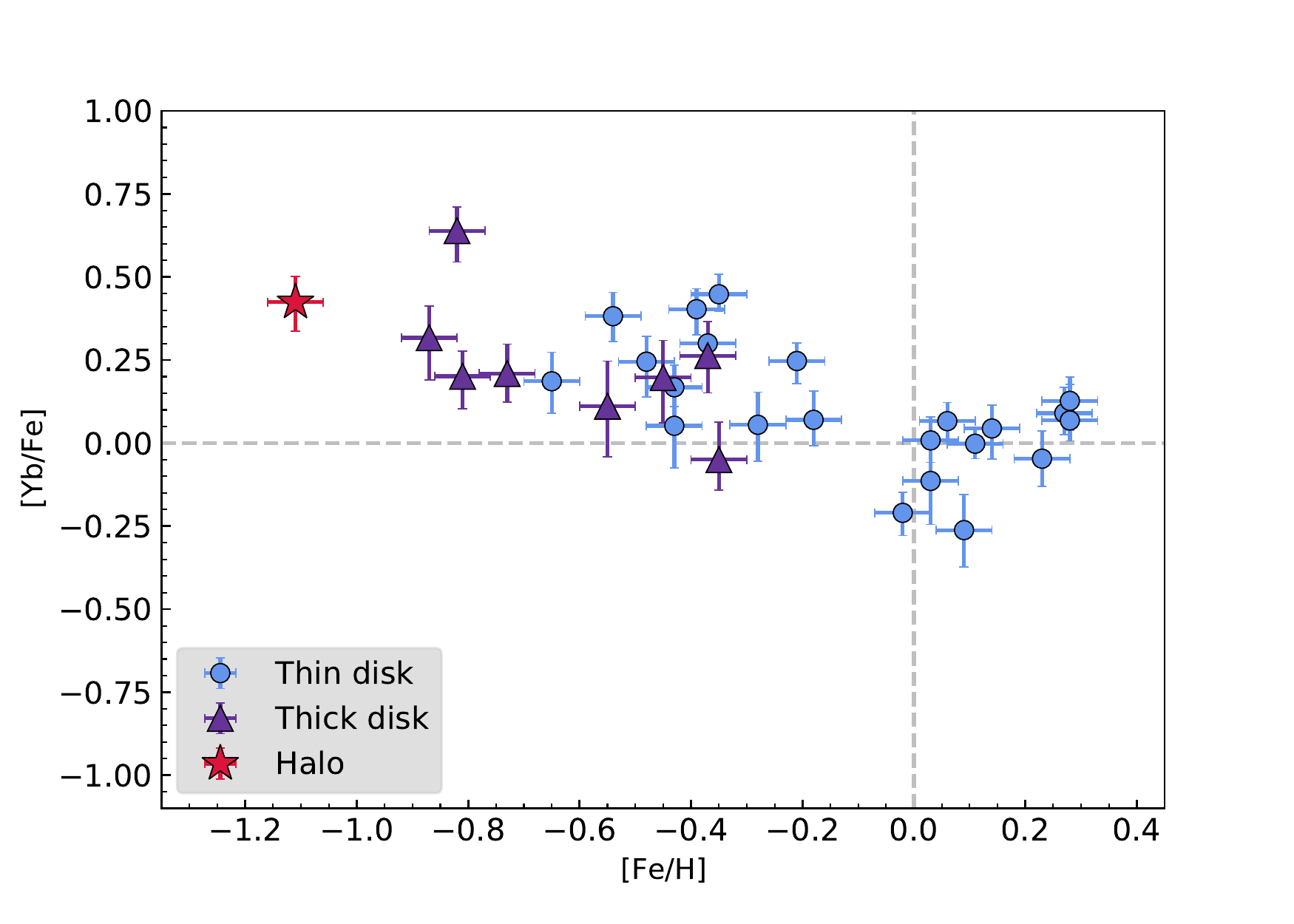}
    \caption{Ytterbium abundances derived in this work. A division into thin disk, thick disk and halo stars has been made based on [Mg/Fe] abundances and the kinematics of the stars, see text. The error bars reflect the first and third quartiles from the Monte Carlo uncertainty estimates presented in Section \ref{sec:unc}, along with a 0.05\,dex uncertainty in [Fe/H]. }
    \label{fig:errorbar}
\end{figure}

The Yb abundances derived in this work are presented in Table \ref{tab:data} and are shown in Figure \ref{fig:errorbar} as [Yb/Fe] versus [Fe/H]. In the figure the error bars are based on the first and third quartiles in the uncertainty calculations described in section \ref{sec:RandUnc} above. 

The spread in abundance that we see in our study is more likely to be the result of uncertainties in the analysis rather than a large cosmic spread in Yb abundances. The error bars represent the uncertainty from stellar parameters only, with factors such as continuum determination and errors in molecular line modelling not being included.

\begin{figure}[h]
\centering
\includegraphics[width=10cm]{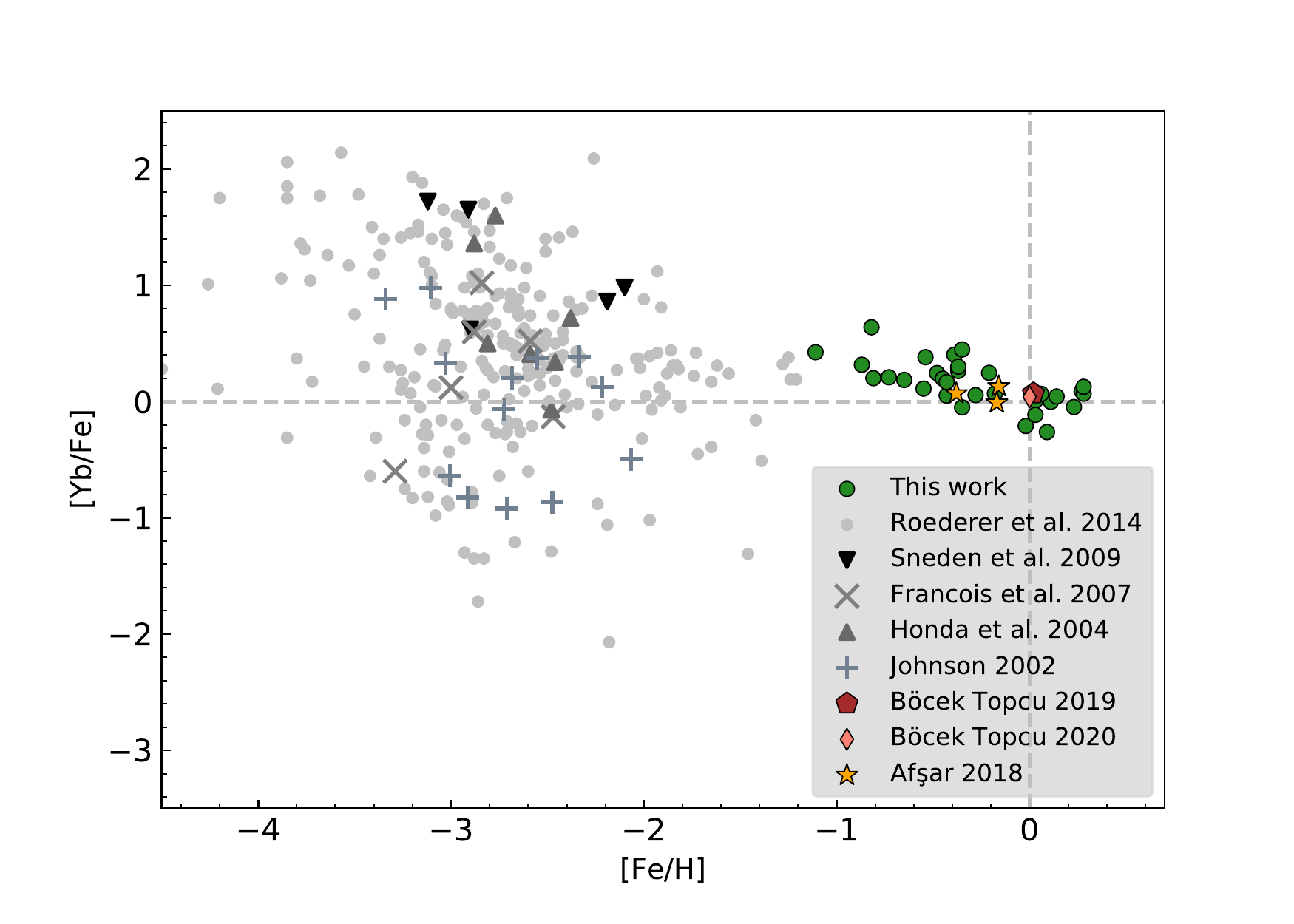}
    \caption{The [Yb/Fe] abundances determined in this work (green) compared to previous work. The studies of halo stars are indicated in various shades of grey: \cite{honda2004ApJ...607..474H}, \cite{johnson2002ApJS..139..219J}, \cite{francois:2007A&A...476..935F}, \cite{roederera2014AJ....147..136R}, \cite{roedererb2014ApJ...791...32R},
    \cite{sneden:2009ApJS..182...80S}. The uncertainties for these works are typically 0.10-0.25\,dex.
    The two open clusters in \citet{bocektopcu2019MNRAS.485.4625B,bocektopcu2020MNRAS.491..544B}, namely NGC 752 and NGC 6940, are indicated in red and pink using their mean abundance, consisting of 10 and 11 stars, respectively. The measured spread in [Yb/Fe] for these clusters is 0.05 and 0.08. The horizontal-branch stars in \cite{afsar2018ApJ...865...44A} are indicted in yellow.}
    \label{fig:AllWork}
\end{figure}

To put these results in context, Figure \ref{fig:AllWork} shows our results together with previously published Yb measurements. The lack of a larger sample of measurements for [Fe/H] $> -1$ illustrates the historic difficulty determining Yb abundances for higher metallicity stars by using the resonance lines in the UV. To validate that similar abundances are obtained with the infrared line used here and the resonance lines is not trivial, since the line we use is weak and thus ill-suited for studying metal-poor stars. 

It may be possible to observe both lines for a portion of the sample in Figure \ref{fig:AllWork}, the very metal-poor stars with [Fe/H] $<$ -2 and [Yb/Fe] $>$ 1. The low metallicity would reduce the blending in the near-UV while the high [Yb/Fe] enhancement would strengthen the H-band line enough to be observable. Such stars typically have a high enhancement in carbon, which, depending on the $\teff$, may cause blending issues in the IR, even for such metal-poor stars. A high resolution and SNR would be required to resolve the line, similar to this study.

The high scatter at lower metallicities, below [Fe/H] $<$ -2, in Figure \ref{fig:AllWork} does not solely originate from the uncertainties, which according to the sources span from 0.1 to 0.25\,dex. It is instead a physical result of inhomogeneities in the Galactic ISM at low metallicity, combined with neutron star mergers and the type of SNe \citep[likely MRSNe][]{kobayashi:2020} hosting the r-process being rare, creating isolated stellar groups of high and low r-abundances \citep[see][and references therein]{sneden2008ARA&A..46..241S}. This behaviour of an r-process dominated element at low metallicities has been reproduced in the stochastic chemical evolution models in \citet{cescutti2015A&A...577A.139C} that takes inhomogeneous mixing into account. As we can see in this study, the scatter decreases substantially at higher metallicities, a result of a more homogeneous stellar disk and regular enrichment.

\subsection{Galactochemical evolution of Yb}
The [Yb/Fe] trend with metallicity shows an enhancement of [Yb/Fe] for stars of subsolar metallicity, decreasing to solar values around solar metallicity, similar to elements produced in SNe type II (for example, the $\alpha$ elements), MRSNe and likely neutron star mergers (r-process) \citep{cescutti2015A&A...577A.139C,grisoni2020MNRAS.492.2828G,kobayashi:2020}. The similarity of $\alpha$- and r-process elemental trends originates from the time scales related to their formation, being a rapid onset of the enrichment in the Galaxy.

At supersolar metallicity the trend appears to flatten, with the exception of the highest metallicity stars. The systematic uncertainties involved in determining abundances for these stars are likely to be high, so we draw no conclusions on the precise slope of the supersolar trend.

To ensure that the Yb-abundances are in line with what is expected for neutron capture elements and to examine the contribution from different production channels, we compare the [Yb/Fe] trend with those of [Ce/Fe] and [Eu/Fe] for the same stars. The Ce abundances are determined from the IGRINS spectra, as described in section \ref{sec: other elem abundances}, while the Eu abundances are from the optical work in \citet{forsberg:2019} which has an 28 stellar overlap with our sample. The very tight abundance trend for Eu indicates a high precision in the analysis. While not using the exact same stellar parameters, they have been determined with a similar method. This should limit systematic uncertainties in the comparison.

\begin{figure*}[h]
\centering
\includegraphics[width=\hsize]{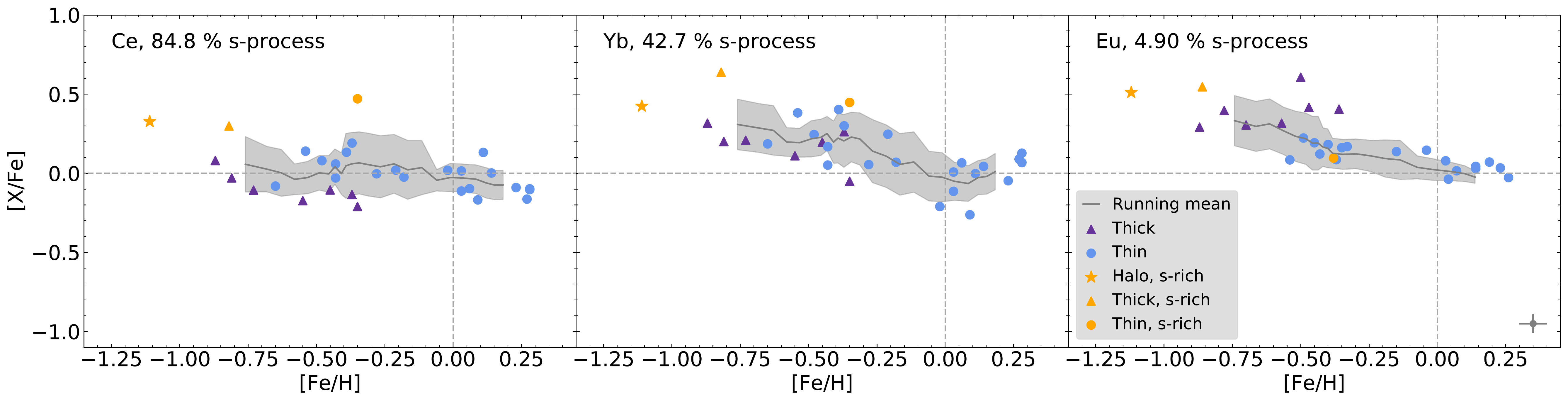}
    \caption{Comparison between the [Ce/Fe] (left) and [Yb/Fe] abundances (middle) derived in this work and [Eu/Fe] (right) abundances from \citet{forsberg:2019}. The stellar populations are indicated by colour and shape, indicated in the legend, where the s-process enhanced stars stand out with yellow colour. The percentage of contribution from the s-process from  \citet{prantzos2020MNRAS.491.1832P} is indicated in the top left corner of each plot. The running mean with a 1$\sigma$ error (grey) have a running-box with a size of 8 stars, roughly 25\% of the sample size. In the lower right corner of the [Eu/Fe] plot the typical uncertainty from \citet{forsberg:2019} is indicated. Dashed grey lines going through [0,0] is the solar value. }
    \label{fig:n-cap}
\end{figure*}

Since Yb is reported to have somewhere between a $40/60\ \%$ to a $50/50\ \%$ contribution from the s- and the r-process, respectively \citep[e.g.][]{birsterzo:2014, kobayashi:2020,prantzos2020MNRAS.491.1832P}, the [Yb/Fe] trend should fit in between the s- and r-process trends of [Ce/Fe] and [Eu/Fe]. In Figure \ref{fig:n-cap}, we plot running means of the full sample, the components of the disk and the s-enhanced stars for the neutron-capture elements. As can be seen, Yb indeed falls nicely in-between the two comparison elements.

In Figure \ref{fig:n-cap} we see that the [Yb/Fe]-trend with metallicity has a similar slope as [Eu/Fe], pointing at an early enrichment from processes with short time scales, such as SNe type II, MRSNe and possibly neutron-star mergers \citep{cescutti2015A&A...577A.139C,grisoni2020MNRAS.492.2828G,kobayashi:2020} \footnote{However, also see the discussion in \citet{skuladottir2020A&A...634L...2S} on the two-folded time-scales of the r-process originating from a joint contribution from SNe and neutron star mergers}.

The [Ce/Fe] trend is much flatter than the [Yb/Fe] trend, although the thick disk s-enhanced star, marked with an yellow triangle in Figure \ref{fig:n-cap} clearly stands out in [Yb/Fe]. The other two s-enhanced stars (halo and thin disk, yellow star and filled circle, respectively) stand out too, but not to the same extent. This is expected since these two do not stand out as much in [Eu/Fe] either, once more pointing at the similarity between these two r-process dominated elements. However, the fact that the s-enriched stars also show a similar pattern in [Yb/Fe] compared to [Ce,Eu/Fe] is a receipt of the precision of determination of the Yb-abundances, and to the smaller extent of s-process contribution to this element.

We can also consider the relative positions of the abundance ratios of the thin- and thick-disk stars in Figure \ref{fig:n-cap}. The thick-disk stars (purple triangles) are lower in the [Ce/Fe] abundance ratios and higher in [Eu/Fe] than the trends of the thin-disk stars (blue filled circles), as expected for s- and r-process dominated elements \citep{forsberg:2019}. The [Yb/Fe] abundance ratios of the thick-disk stars are also found to be {\it lower}, although only slightly, than those of the thin-disk stars. This is a clear indication of the s-process contribution to this element. At the same time, the shape of the overall [Yb/Fe] trend, seen from the running mean, gives clear indication of the r-process contribution, as expected from Galactic chemical evolution models of this ratio. Nonetheless, \citet{kobayashi:2020} report Yb to "have a significant contribution from the s-process".

\begin{figure}[h]
\centering
\includegraphics[width=8cm]{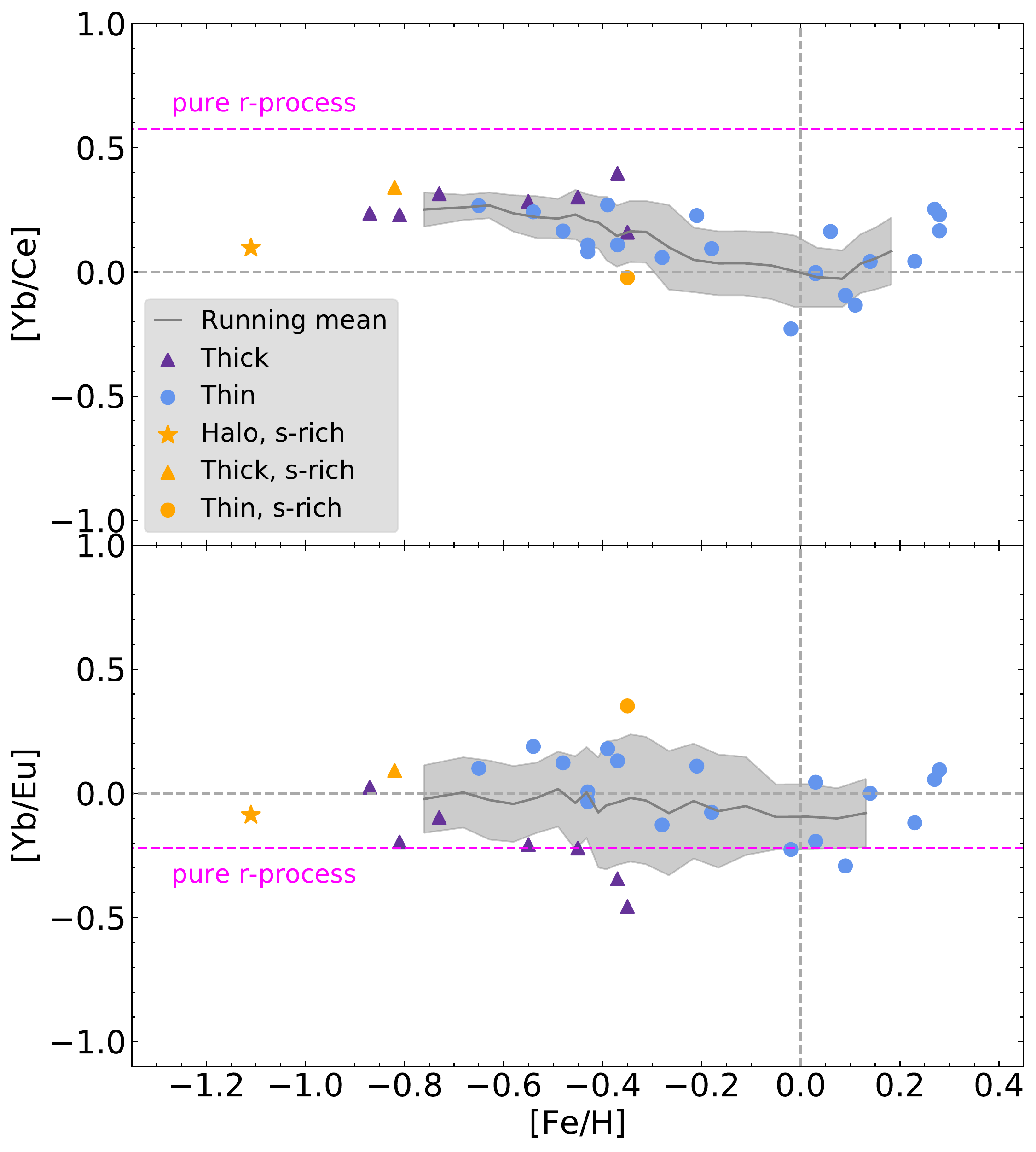}
    \caption{[Yb/Ce] (top) and [Yb/Eu] (bottom) plotted towards [Fe/H]. Similar notation as in Figure \ref{fig:n-cap}, and [Eu/Fe] abundances originating from \citet{forsberg:2019}. The pure-r-process (solar) contribution, indicated in dashed magenta, are calculated using the values in \citet{prantzos2020MNRAS.491.1832P}, as indicated in Figure \ref{fig:n-cap}.}
    \label{fig:pure r-process}
\end{figure}

To further investigate the contribution from the s- and r-process, we plot [Yb/Ce] and [Yb/Eu] versus metallicity, see Figure \ref{fig:pure r-process}. A flat trend in these type of plots indicates a similar production rate of the two elements, whereas a decrease/increase on the other hand indicates discrepancies.

The so-called "pure r-process line" is also indicated in the plot, which is calculated using the solar r-process contributions of the elements. As such, the pure r-process is the value of the r-process contribution in both elements, such that [r-process(Yb)/r-process(Ce)], for instance. The closer to the pure-r-process line, the more of the elements originate from the r-process.

Considering [Yb/Ce] in Figure \ref{fig:pure r-process}, it becomes clear that the r-process component is stronger in Yb than in Ce. The r-process dominates the production of neutron-capture elements at lower metallicities, also producing the s-process dominated elements, like Ce, and Yb to a higher extent. The onset of s-production in AGB stars, originating from low- to intermediate-mass stars, has a time delay in enriching the Galactic ISM. This can be seen at around [Fe/H] $\sim$ -0.3 where the [Yb/Ce] trend starts decreasing, due to an increase of s-process production and a relatively higher Ce enrichment compared to Yb.

The [Yb/Eu] trend clearly indicates a significant contribution from the r-process in the production of Yb, similar to what is seen when comparing the trends in Figure \ref{fig:n-cap}. We also note that the thick-disk stars tend to all lie around the pure r-process line, indicating a large r-process contribution in the thick disk.

Because of potential issues with finding local continuum points and the risk of stronger molecular blends for supersolar metallicities, we refrain from drawing any conclusion about the possible upturns we see in this metallicity region. 

The above comparisons with other neutron-capture elements strengthen the validity of the Yb abundances presented in this work, and assures us that we can correctly model the CO-blend in the Yb\,II-line.

\section{Conclusions\label{sec:conclusion}}
In this work we have presented Yb abundances for 30 K-giants, with metallicities in the -1.1 $<$ [Fe/H] $<$ 0.3 range, which to the best extent of our knowledge, is the largest disk sample with Yb abundances to date. Our typical (random) uncertainties in [Yb/Fe] are approximately 0.1\,dex.

The derived abundances align well with previous studies of low metallicity stars. Although the Yb\,II line is not useful in the solar spectrum at this resolution, the measured trend obtained from abundance determinations in our K-giants passes through the solar value at [0,0], which is reassuring. By comparing with abundances of two other neutron-capture elements, namely Ce and Eu, we find the cosmic origin of Yb to be dominated by the r-process, supported by the [Yb/Eu] comparison. It is, however, clear from the s-enhanced stars and the precise alignment between the thin and thick disk that the s-process plays a part in producing Yb, as expected from theoretical models. Additionally, we find the Yb-abundances to be of high quality since they reproduce the s-enhancement previously observed for the same stars in optical spectra, confirming that the CO-blend in the Yb-II line is modelled properly. 

Previous measurements of neutron-capture elements in stars from the infrared H- and K-bands are dominated by Ce, which has a number of usable lines \citep{cunha:2017}. Two other elements have been measured in small samples, Yb and Nd \citep{hasselquist:2016}. Like Ce, Nd is thought to be produced predominately in the s-process for stars of solar metallicity \citep[see e.g.][]{kobayashi:2020}, but at a higher uncertainty \citep{bocektopcu2019MNRAS.485.4625B,bocektopcu2020MNRAS.491..544B}.

Elements created in the r-process offer clear signatures of events leading to element formation on short time scales, such as neutron star mergers. The ability to determine abundances of the r-process dominated element Yb from near-infrared spectra for a wide range of metallicities up to supersolar values, opens an additional Galactic chemical evolution channel from near-IR spectra. This is significant both for the readily available near-IR spectrographs and upcoming versatile instruments, such as the HIRES spectrograph for the ELT \citep{HIRES:2018}. 

On the usefulness of having a wide range of neutron capture elements in Galactochemical research, we can consider our comparison of Yb to Ce and Eu and it becomes evident that the thick disk has a stronger enrichment by the r-process, compared to the thin disk. The vice-versa holds for the thin disk seemingly enriched in s-process, compared to the thick disk.

We have shown that with high enough spectral resolution and a careful analysis, the territory of the r-process can thus now be reached in the near-infrared. This will help to unravel regions previously obscured by dust, such as the Milky Way bulge. For future large near-IR spectroscopic surveys the Yb\,II line could therefore offer the study of the r-process also in obscured stellar populations. Here, Yb can contribute a lot in deciphering the star formation history and assembly of the Bulge.

\begin{acknowledgements}
M.M. acknowledges funding through VIDI grant "Pushing Galactic Archaeology to its limits" VI.Vidi.193.093, which is funded by the Dutch Research Council (NWO). R.F's and A.J's research is supported by the Göran Gustafsson Foundation for Research in Natural Sciences and Medicine. R.F. and N.R. acknowledge support from the Royal Physiographic Society in Lund through the Stiftelse Walter Gyllenbergs fond and Märta och Erik Holmbergs donation. This work used the Immersion Grating Infrared Spectrometer (IGRINS) that was developed under a collaboration between the University of Texas at Austin and the Korea Astronomy and Space Science Institute (KASI) with the financial support of the Mt. Cuba Astronomical Foundation, of the US National Science Foundation under grants AST-1229522 and AST-1702267, of the McDonald Observatory of the University of Texas at Austin, of the Korean GMT Project of KASI, and Gemini Observatory.

These results made use of the Lowell Discovery Telescope (LDT) at Lowell Observatory. Lowell is a private, non-profit institution dedicated to astrophysical research and public appreciation of astronomy and operates the LDT in partnership with Boston University, the University of Maryland, the University of Toledo, Northern Arizona University and Yale University.

This paper includes data taken at The McDonald Observatory of The University of Texas at Austin. 

This work has made use of the VALD database, operated at Uppsala University, the Institute of Astronomy RAS in Moscow, and the University of Vienna.
\end{acknowledgements}

\bibliographystyle{aa} 
\bibliography{references.bib} 

\end{document}